\documentclass[aps,twocolumn,prb,preprintnumbers,amsmath,amssymb,superscriptaddress,floats]{revtex4-2}
\usepackage{graphicx}
\usepackage{bm}
\usepackage{natbib}
\usepackage{float}
\usepackage{multirow}
\usepackage{array}
\usepackage{physics}
\usepackage{siunitx}
\usepackage{verbatim}
\usepackage{xcolor}
\usepackage{amsmath,lipsum}

\def\pzo{Pr$_2$Zr$_2$O$_7$}

\def\pzo{Pr$_2$Zr$_2$O$_7$}
\def\nzo{Nd$_2$Zr$_2$O$_7$}
\def\pio{Pr$_2$Ir$_2$O$_7$}
\def\nio{Nd$_2$Ir$_2$O$_7$}

\def\cm{cm$^{-1}$}

\begin{document}

\title{Scattering of Raman  phonons on magnetic and electronic excitations in  pyrochlores Nd$_2$Zr$_2$O$_7$ and Nd$_2$Ir$_2$O$_7$.}

\author{Sami Muhammad}
\affiliation{Institute for Quantum Matter and Department of Physics and Astronomy, Johns Hopkins University, Baltimore, Maryland 21218, USA}

\author{Yuanyuan Xu}
\affiliation{Institute for Quantum Matter and Department of Physics and Astronomy, Johns Hopkins University, Baltimore, Maryland 21218, USA}

\author{Christos Kakogiannis}
\affiliation{Institute for Quantum Matter and Department of Physics and Astronomy, Johns Hopkins University, Baltimore, Maryland 21218, USA}

\author{Takumi Ohtsuki}
\affiliation{Institute for Solid State Physics, University of Tokyo, Kashiwa, Chiba 277-8581, Japan}

\author{Yang Qiu}
\affiliation{Institute for Solid State Physics, University of Tokyo, Kashiwa, Chiba 277-8581, Japan}

\author{Satoru Nakatsuji}
\affiliation{Institute for Solid State Physics, University of Tokyo, Kashiwa, Chiba 277-8581, Japan}
\affiliation{Department of Physics, University of Tokyo, Bunkyo-ku, Tokyo 113-0033, Japan}
\affiliation{Institute for Quantum Matter and Department of Physics and Astronomy, Johns Hopkins University, Baltimore, Maryland 21218, USA}
\affiliation{CREST, Japan Science and Technology Agency, Kawaguchi, Saitama 332-0012, Japan}
\affiliation{Trans-scale Quantum Science Institute, University of Tokyo, Bunkyo-ku, Tokyo 113-0033, Japan}

\author{Eli Zoghlin}
\affiliation{Materials Department, University of California, Santa Barbara, Santa Barbara, California 93427, USA}
\affiliation{Institute for Quantum Matter and Department of Physics and Astronomy, Johns Hopkins University, Baltimore, Maryland 21218, USA}

\author{Stephen D. Wilson}
\affiliation{Materials Department, University of California, Santa Barbara, Santa Barbara, California 93427, USA}

\author{Natalia Drichko}
\affiliation{Institute for Quantum Matter and Department of Physics and Astronomy, Johns Hopkins University, Baltimore, Maryland 21218, USA}
\email{Corresponding author. Email:drichko@jhu.edu}

\begin{abstract}
Magnetic rare earth atoms on pyrochlore lattices can produce exotic magnetic states such as spin ice and quantum spin ice. These states are a result of the frustration in the pyrochlore lattice, as well as crystal field degrees of freedom of rare earth atoms, and their interactions with the lattice. Raman scattering spectroscopy, which possesses high spectral resolution and can easily access broad energy and temperature ranges, is an optimal tool to study these excitations and their interactions. In this work we follow the Raman scattering of zone center phonons and crystal field excitations of Nd$^{3+}$ in \nzo\ and \nio\ in the temperature range where these materials are paramagnetic. A comparison between an insulating \nzo\ and a semimetallic  \nio\  allows us to distinguish between scattering of phonons on other phonons, crystal field excitations, and electrons, highlighting interactions between these degrees of freedom.
\end{abstract}

\date{\today}
\maketitle

\section{Introduction}

Frustrated 3D lattices of pyrochlores of a general formula A$_2$B$_2$O$_7$ with rare earth atoms at the A site are known to host exotic magnetic states such as spin ice, quantum spin ice  states, and all-in-all-out  (AIAO)  order~\cite{Gardner2010,Gingras2014,Rau2019}.  This research is still in its active phase, with insulating pyrochlores being the most understood. Recently, a new hope of finding a quantum spin liquid  state appeared with the discovery of this state in a pyrochlore Ce$_2$Zr$_2$O$_7$~\cite{Gaudet_CZO_2019,Gao_CZO_2019,Smith_CZO_2022,Bhardwaj_CZO_2022}. One of the compounds studied in this work, \nzo, shows a quantum spin-ice phase in the temperature regime above the ordering into an AIAO state below T$_N \approx$ 0.3~K~\cite{Lhotel2021,Xu_NZO_2020}. 

While materials with Zr on the B site are insulating, if a B site is occupied by Ir, its extended orbitals lead to semimetallic properties. A symmetry-protected quadratic band touching at the $\Gamma$-point has drawn attention to these materials in the first place~\cite{Witczak2012}. Since Ir is magnetic with $J$=1/2 and has dominant antiferromagnetic interactions, Ir moments order at T$_N^{Nd}$=33~K in an AIAO order~\cite{Matsuhira2011,Ueda2012}.  This order breaks time reversal symmetry and can lead to a Weyl semimetal state~\cite{Wan2011,Witczak2014,Savary2016}, as was recently confirmed experimentally~\cite{Nikolic2024}. A combination of two magnetic sublattices, a rare earth on an A site and Ir on a B site, and an interplay or competition of various magnetic interactions leads to exotic magnetism of the rare earth at elevated temperatures compared to the insulating pyrochlores, such as quantum-spin ice of Ho moments in Ho$_2$Ir$_2$O$_7$~\cite{Pearce2022}   and proposed spin ice of Nd moments in   \nio\ at intermediate temperatures~\cite{Xu2024}, while the ground state is an AIAO order.

Understanding of interactions of the lattice with magnetic and electronic degrees of freedom in pyrochlores is an important way to fully understand the origin of the exotic magnetic and electronic states in these materials. A broad range of experimental and theoretical evidence of interactions of the lattice with magnetic degrees of freedom already exists, with those relevant to the studied compounds listed below. For rare-earth pyrochlores, interactions of the lattice with the dipole-octopole degree of freedom are exemplified  experimentally and theoretically in \pzo~\cite{Tang2023}. Modulation of crystal field excitations by phonons, so called vibronic coupling, is frequently observed in rare earth pyrochlores~\cite{Xu2021cef,Gaudet2018,Thalmeier1982}. Exotic spin- phonon coupling mediated by spin-orbit coupling is suggested in pyrochlores with 5d atoms~\cite{Sohn_COO_2017,Son_YIO_2019}. Especially relevant to this study is the theoretical work which demonstrates how phonon Raman scattering  can indirectly probe quantum  spin-ice excitations~\cite{Seth2022}. Raman scattering spectroscopy,  which probes $\Gamma$-point phonons with high spectral resolution, is a unique method to access this kind of information.

In this work we follow the temperature dependence of lattice phonons and crystal field excitations of two Nd-based pyrochlores, an insulating \nzo\ and a semimetallic \nio, in the  paramagnetic state. For \nzo\ we follow these excitations down to 10~K.  For \nio\ we follow them above the temperature of AIAO ordering of Ir moments, T$^{Ir}_N$=33~K, while the data in the magnetically ordered state are presented elsewhere~\cite{Xu2024}. While this temperature range is above the temperatures of magnetic and electronic states which attracted the most attention to these materials, it provides an important insight into the interactions of the lattice with electronic excitations and crystal field excitations.


\section{Experimental}

A \nio~single crystal was grown by the KF-flux method~\cite{Millican2007} and it has an as-grown
octahedron-shaped (111) facet which was used for the measurements. A \nzo\ single crystal was grown using the floating zone technique~\cite{Zoghlin2021}.

Raman scattering spectra were collected using the Jobin-Yvon T64000 triple monochromator spectrometer equipped with a liquid nitrogen cooled CCD detector. Measurements were performed in a back-scattering geometry 
using a micro-Raman channel equipped with an Olympus microscope, using a laser probe of 2$\mu$m in diameter.   A 514.5 nm line of Ar$^+$-Kr$^+$ mixed gas laser was used as the excitation light. 

For low temperature measurements, the samples were mounted on the cold-finger of a Janis ST-500 cryostat, which can be cooled down to 4K without laser heating. The laser power was kept at 0.5~mW for the measurements of \nzo. The semimetallic \nio\ has at least one order of magnitude lower Raman scattering signal. Therefore a laser power of 2 mW was used for these measurements, which resulted in heating of the sample by about 20~K. The temperatures presented in the paper are corrected for the laser heating.

For the measurements, the large floating-zone-grown crystals of \nzo\ were oriented and cut, so that the Raman scattering measurements were taken from the (001) plane with the polarization vector of the excitation laser $e  \parallel a$, where $a$ is the crystallographic axis. Measurements were performed for  $(x,x)$ and $(x,y)$ scattering channels. Measurements were done in the spectral range between 10-90 meV.

\nio\ flux-grown crystals typically have a pyramid shape with each direction less than 1 mm and naturally grown (111) facets. Raman scattering measurements were done from a cleaved surface of the (111) facet of the crystal in the $(x,x+y)$ scattering channel. Data in the spectral range from ~10-90 meV were taken in the temperature range from 40 to 270~K. In order to obtain a temperature dependence for the phonons of \nio\ in the broad temperature range despite having a very low signal, the data were taken on continuous heating of the sample with a heating rate of 0.1~K/min. The data were averaged over 1 hour, which corresponds to averaging over 6~K. The averaging temperature range was selected in a way that the change in the spectra of the two successive averaging temperature ranges $\Delta T_1$ and $\Delta T_2$ was smaller than the noise amplitude A: I($\Delta T_1$)-I($\Delta T_2$) $<$ A. 

All spectra were normalized using the Bose-Einstein thermal population factor $[n(\omega)+1]$, where $n(\omega)$ is the Bose occupation factor.

In order to obtain the temperature dependence of the phonon parameters demonstrated in Fig.~\ref{Combined_Parameters}, the phonon spectra were fit by a sum of Voigt lineshapes, where the Gaussian component width was kept at 1.5~\cm\ in order to correct for the resolution of the spectrometer.  
The micro-channel Raman scattering measurements are known to introduce an artifact background for highly reflective samples with weak Raman response, which is true for \nio. Such artifact background exceeds in amplitude electronic scattering which can be detected for this material in pseudo-Brewsters geometry~\cite{Nikolic2024}.  To correctly analyze the Raman phonon scattering of \nio, which is the focus of this paper, we have identified the temperature-independent artifact background and subtracted it from all the data as a part of the primary data analysis. This background was identified in the following procedure: A spectrum at a given temperature was fit by a sum of Voigt lineshapes
associated with the sharper features of phonons, and Gaussian lineshapes representing the background.  
Next, all Voigt fits were subtracted from the raw data to estimate a background. This background was fit to several new temperature independent Gaussians and a temperature dependent linear background. This total background was then subtracted from the raw spectra at each temperature to provide a phonon spectrum analyzed in this manuscript.


\section{Results}

\begin{figure}
  \centering
  \includegraphics[width=0.5\textwidth]{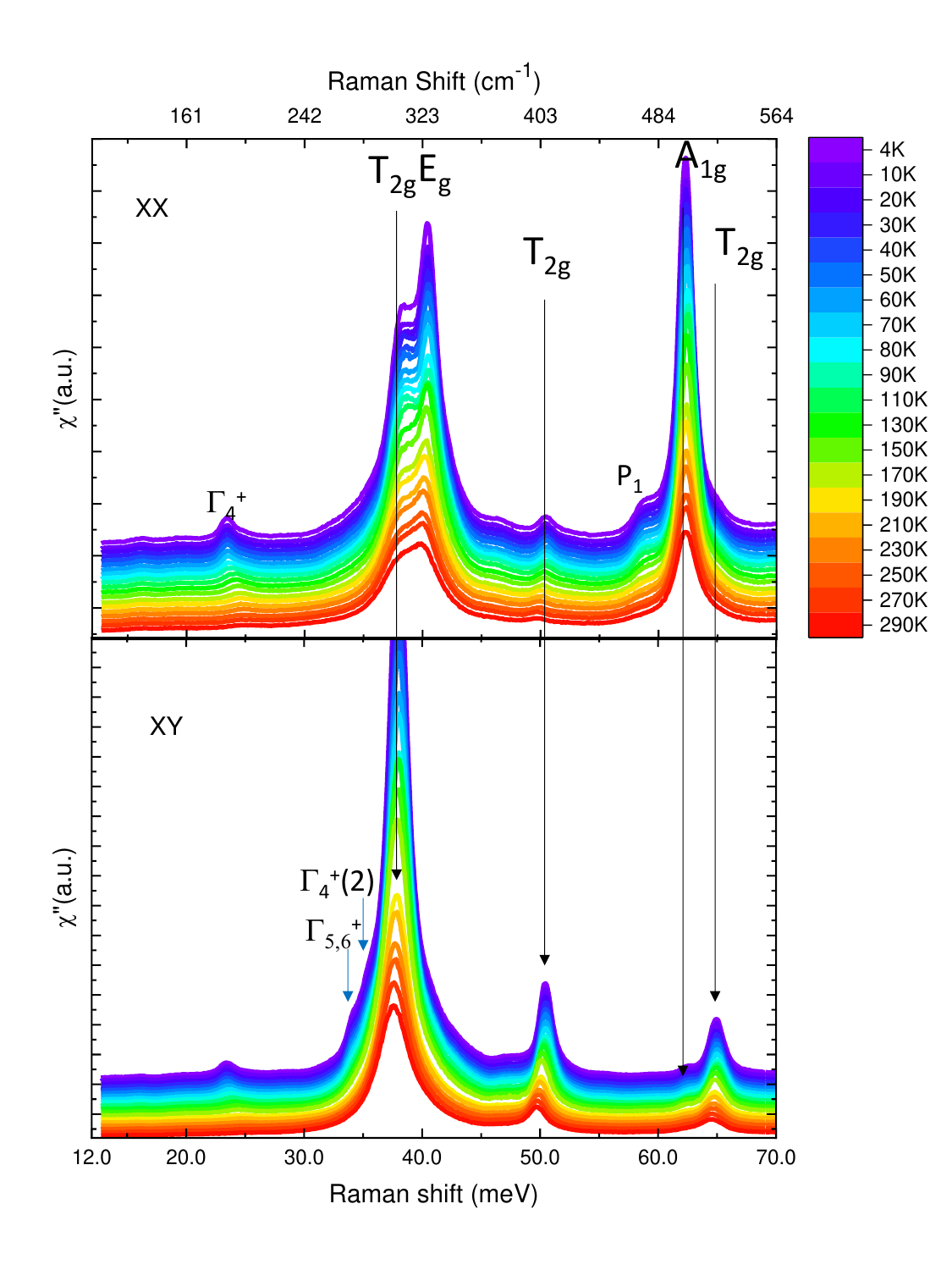}
  \caption{Raman scattering spectra of \nzo\ at temperatures between 290 and 15 K, upper panel shows $(x,x)$ scattering channel, lower panel shows $(x,y)$ scattering channel. \label{NZO_Raman}}
\end{figure}

Figs.~\ref{NZO_Raman} and \ref{NIO_Raman} present temperature dependent Raman scattering spectra of \nzo\ in $(x,x)$ and $(x,y)$ scattering channels in 12- 70 meV spectral range and \nio\ in  $(x,x+y)$ in 35- 70 meV spectral range, where we expect to observe phonons and crystal field excitations within the ground state $^4I_{9/2}$ multiplet. Dependence of the spectra on temperature and polarization allows us to distinguish between phonons and crystal field excitations. Typically, sharp peaks observed at room temperature are the excitations of $\Gamma$-point phonons, while crystal electric field (CEF) excitations appear in the spectra as well-defined peaks as the temperature is lowered due to larger scattering at higher temperatures~\cite{Cardona2000}.

\subsection{Phonon Raman scattering}

Both \nzo\ and \nio\ crystal structures belong to the   $Fd\bar{3}m$ (No.~227) space group, which corresponds to the $O_h$ point group.
Wyckoff positions and respective Raman activity of the atoms are presented in Table~\ref{table:wyckoff}.  This demonstrates that all the Raman-active phonons in these materials are related to the oxygen motion. Phonons were assigned using polarization dependence as noted in Table~\ref{table:intensity} and following DFT calculations of phonons for \pzo\ and \pio\ presented in ~\cite{Xu2021cef,Xu2022phonons}. 
 
\begin{table}[H]
	\centering
	\caption{Wyckoff positions and $\Gamma-$point representations for \nzo\ and \nio.\label{table:wyckoff}}
	\begin{ruledtabular}
	\begin{tabular}{
		>{\raggedright\arraybackslash}p{0.2\linewidth}
		>{\centering\arraybackslash}p{0.3\linewidth}
		>{\centering\arraybackslash}p{0.4\linewidth}}
		Element & Wyckoff positon & $\Gamma$ representation \\
		\hline
		Nd & $16c$ & Inactive \\
		Zr/Ir & $16d$ & Inactive \\
		O & $48f$ & $A_{1g} + E_g + 3T_{2g}$ \\
		O$'$ & $8a$ & $T_{2g}$ \\
	\end{tabular}
	\end{ruledtabular}
\end{table}

\begin{table}[H]	
	\centering
	\caption{Components of Raman tensor for $(x,x)$ and $(x,y)$ polarizations.\label{table:intensity}}
	\begin{ruledtabular}
	\begin{tabular}{
		>{\raggedright\arraybackslash}p{0.2\linewidth}
		>{\centering\arraybackslash}p{0.2\linewidth}
		>{\centering\arraybackslash}p{0.2\linewidth}
		>{\centering\arraybackslash}p{0.2\linewidth}}

		Geometry & $A_{1g}$ & $E_{g}$ & $T_{2g}$\\
		\hline
		$(x,x)$ & $a^2$ & $b^2$ & $c^2$\\
		$(x,y)$ & 0 & $b^2$ & $\frac{2}{3}c^2$\\
	\end{tabular}
	\end{ruledtabular}
\end{table}


\subsubsection{Phonons of \nzo}

\begin{figure}
  \centering
  \includegraphics[width=0.5\textwidth]{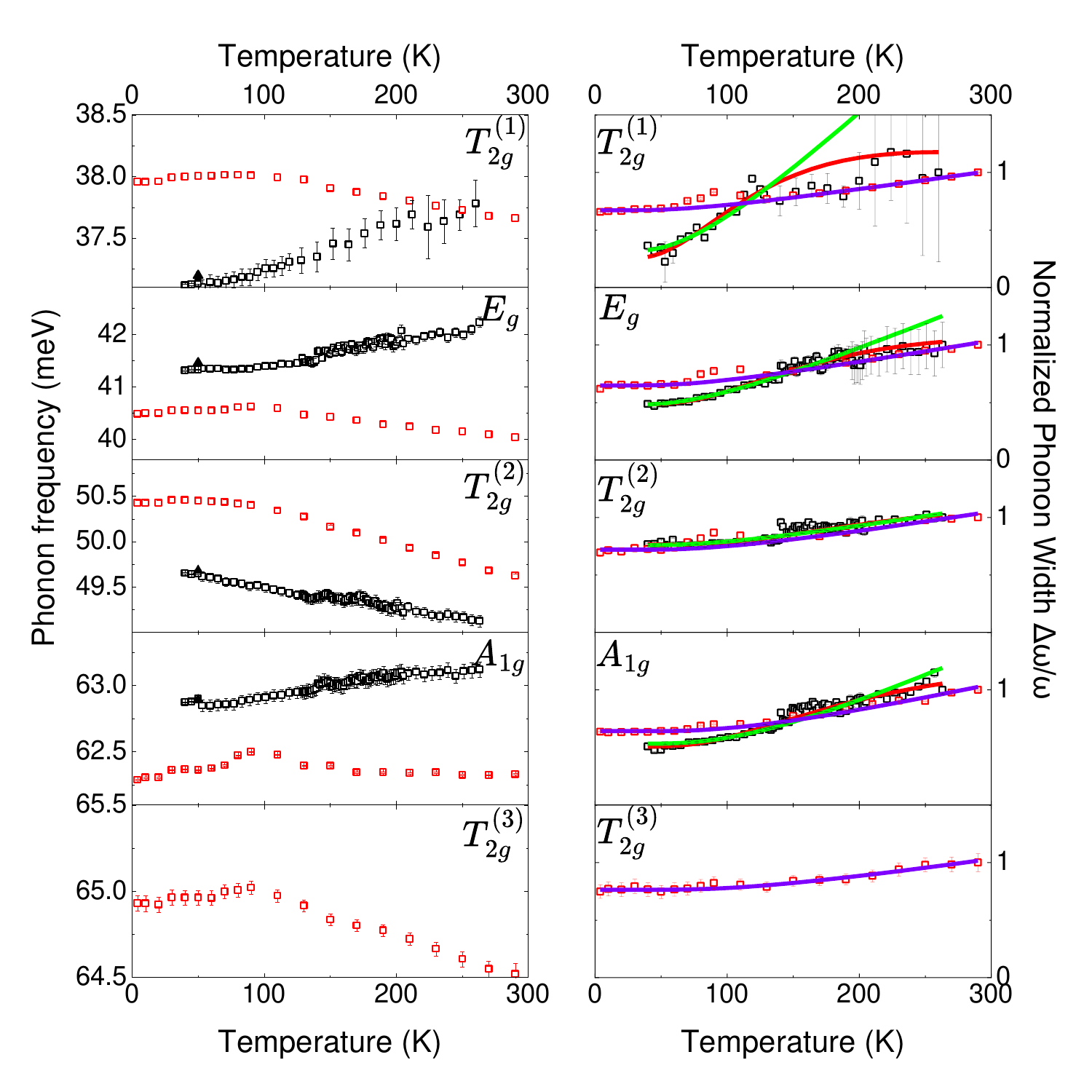}
  \caption{Temperature dependence of frequencies (left panel) and width (right panel) of Raman active phonons in \nzo (red squares) and \nio (black squares). 
  In the right panel, solid lines show temperature dependence of phonon line width of \nzo (violet) and \nio (green) suggested by  the Klemens model scattering. We also show electron-phonon models for \nio (red). Note that we present phonon width normalized on the width at high temperatures ($\gamma(T)/\gamma(290~K)$) in order to have a possibility to compare the temperature dependence of the phonons of the two materials.}\label{Combined_Parameters}
\end{figure}

Temperature dependent Raman phonon spectra of \nzo\ in $(x,x)$ and $(x,y)$ scattering channels are shown in Fig. 1. Phonons are assigned based on their polarization dependence, see Table~\ref{table:intensity}. Their frequencies together with their assigned symmetries are listed in Table~\ref{table:nzo phonons}. T$^{(1)}_{2g}$ and E$_g$ phonons have very close frequencies, but can be distinguished due to different intensities in the $(x,x)$ and $(x,y)$ scattering channels. Parameters of the phonons show weak temperature dependence (Fig.~\ref{Combined_Parameters}). Crystal field excitations marked in Fig~\ref{NZO_Raman} will be discussed in Sect.~\ref{CF}. The feature marked as P1 at 472~\cm\ appears in the spectra in both scattering channels. To the best of our knowledge, it cannot be assigned to Raman active phonons nor to crystal field excitations calculated using a 
point charge  approximation with D$_{3d}$ symmetry of the Nd$^{3+}$ site~\cite{Watahiki2011,Xu2015}.


\begin{table}
\caption{Phonon frequencies of \nzo\ and \nio\ with the assignment based on the polarization dependence and DFT calculations for \pzo\ and \pio\ presented in ~\cite{Xu2022phonons}} 
\begin{tabular}{ c|c|c }
  \hline
 \nzo (meV) & \nio (meV) & Symmetry \\ \hline
  37.9 & 37.1  &  T$^{(1)}_{2g}$ \\
  40.5 &  41.4 &   E$_g$\\
  50.4 &  49.6 &  T$^{(2)}_{2g}$  \\
  58.5 w &   &   P1 \\
  62.3 & 62.8  &  A$_{1g}$ \\
  64.9 & 68.5  &  T$^{(3)}_{2g}$\\
  \hline
\end{tabular}
\label{table:nzo phonons}
\end{table}

Temperature dependent parameters of the \nzo\ phonons obtained from the fit of the spectra by a sum of Voigt shapes are presented in Fig.~\ref{Combined_Parameters} (red color points). The overall hardening and narrowing  of the phonons on cooling follows the standard behavior determined by the contraction of the unit cell and the decrease of phonon scattering. In the absence of  scattering other than phonon-phonon, the temperature dependence of scattering $\gamma$ of $\Gamma$-point optical phonons typically follows the Klemens model~\cite{Kim2012}:
 $\gamma(T,\omega)=\gamma_0+A(2n_B(\omega/2)+1)$,  where $\gamma_0$ is a temperature independent term determined by disorder, $\omega$ is the phonon frequency, and $n_B$ is Bose-Einstein statistical factor determining levels population at temperature T for $\omega/2$, $n_B(\omega/2)=(e^{\frac{\omega h}{2kT}}-1)^{-1}$. In Fig~\ref{Combined_Parameters} (right panel), solid lines show a normalized on high temperatures phonon linewidth $\gamma(T)/\gamma(290~K)$ (red squares) together with the fit by the Klemens model (violet line). While the overall temperature dependence of phonon width follows the Klemens model, for  all the phonons we observe a deviation from it at about 100~K, where the width exceeds the expected value, but then decreases again and follows the conventional curve. Frequencies of the phonons also indicate 100~K is a characteristic temperature. The hardening for all the phonons is observed only down to 100~K. Below this temperature, the dependence of phonon frequency for phonons at around 38 and 50 meV flattens, while the other three phonons soften on further cooling. This behavior demonstrates that some other factor beyond regular phonon-phonon scattering is important for the lattice response of \nzo.



\begin{table}
\caption{Klemens model fit parameters for \nzo, according to FWHM $\gamma(T,\omega)=\gamma_0+A(2n_B(\omega/2)+1)$. 
 }
\begin{tabular}{ c|c|c|c|c }
  \hline
  $\text{Phonon}$ & Polarization & $\omega \text{(meV)}$ & A \text{(meV)} & $\gamma_{0} \text{(meV)}$ \\ \hline
  T$_{2g}^{(1)}$ & XY & 37.9  & 0.6 & 1.7 \\
  $E_{g}$ & XX& 40.5 & 0.8 & 1.3\\
  $T_{2g}^{(2)}$ &XY&  50.4 & 0.6 & 1.0 \\
  $A_{1g}$ & XX& 62.3 & 1.1 & 0.4  \\
  $T_{2g}^{(3)}$ &XY& 64.9  &  0.8 & 1.0 \\
  \hline
\end{tabular}
\label{table:nzo_fits}
\end{table}


\begin{table}
\caption{Electron-phonon model fit parameters for \nio, according to FWHM $\gamma(T,\omega)=\gamma_0+F(n_F(\omega_a)+n_F(\omega_a + \omega))$.
}
\begin{tabular}{ c|c|c|c|c }
  \hline
  $\text{Phonon}$ & $\omega \text{(meV)}$ & F $\text{(meV)}$ & $\gamma_{0} \text{(meV)}$ & $\omega_a \text{(meV)}$ \\ \hline
  T$_{2g}^{(1)}$ & 37.1 & 4.7 & 0.3 & 18.8 \\
  E$_{g}$ & 41.4 & 9.2 & 1.5 & 27.8\\
  T$_{2g}^{(2)}$ &  49.6 & 4.2 & 1.6 & 39.2 \\
  A$_{1g}$ & 62.8 & 4.3 & 0.8 & 30.6  \\
  T$_{2g}^{(3)}$ &  &  &  \\
  \hline
\end{tabular}
\label{table:nio_fits}
\end{table}

\subsubsection{Phonons of \nio}

\begin{figure}
  \centering
  \includegraphics[width=0.5\textwidth]{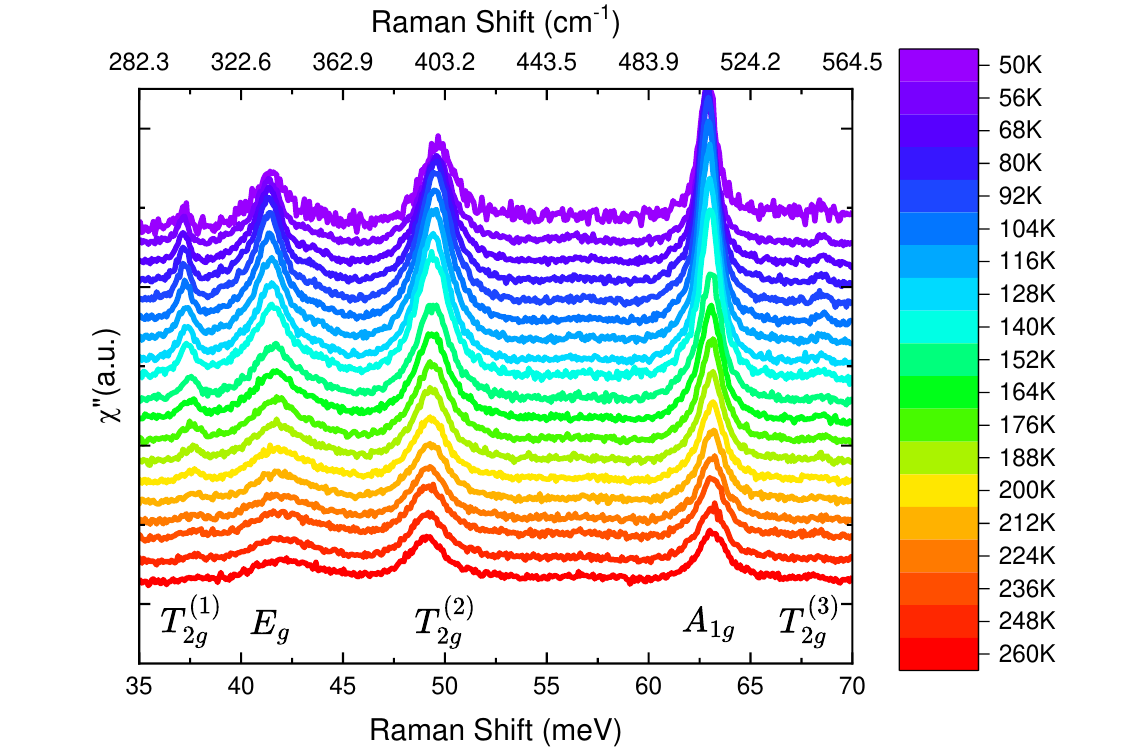}
  \caption{Raman spectra of \nio\ between 50 and 260 K in (x,x+y) polarization }\label{NIO_Raman}
\end{figure}

Fig.~\ref{NIO_Raman} shows temperature dependent Raman spectra of \nio\ in 35-70 meV spectral range in $(x,x+y)$ scattering channel in the paramagnetic state from room temperature down to 50K. Since the structure of \nio\ is  very close to that of \nzo\ we expect phonons of the same symmetries and similar energies, despite the overall intensity of the spectra about two orderes of magnitude lower due to the semimetallic nature of the material. The comparison of frequencies of oxygen phonons for these two materials and their assignment is presented in Table~\ref{table:nzo phonons}. We note that the third T$_{2g}$ mode can be observed as a small peak just below 70~meV, but the intensity is too weak for a discussion of its behavior beyond qualitative softening and narrowing upon cooling.

Temperature dependent parameters of the \nio\ phonons are presented in Fig.~\ref{Combined_Parameters}. We can see that in contrast to \nzo, the observed  phonons soften upon cooling, except for the second T$_{2g}$ mode which hardens. The phonons which lie above approximately 45 meV (E$_g$ and T$_{2g}^{(2)}$)  narrow on lowering the temperature in the paramagnetic state above T$_N^{Ir}$ with the dependence which is in a reasonable agreement with the  Klemens model as shown by the red fitting curves in Fig.~\ref{Combined_Parameters}.  Two lower frequency phonons show a decrease of the linewidth with temperature much faster than the same phonons in \nzo, and the description of the decrease of scattering on cooling by Klemens model fails. This energy range overlaps with the electronic excitations in the semimetallic bands~\cite{Nikolic2024}.  Indeed, the temperature dependence of these phonons is described  by the scattering of phonons on interband transitions $\Gamma_{\mathrm{ph-el}}(T) = \Gamma_0 + F(n_F(\hbar\omega_a, T) - n_F(\hbar\omega_a + \hbar\omega_{\rm{ph}}, T))$, as observed in other semimetals, including \pio~\cite{Osterhoudt2021,Xu2022phonons}, the parameters of the fit are shown in Table~\ref{table:nio_fits}. 

The details of the behavior of the phonons in the magnetically ordered state below T$_N^{Ir}$=33~K, where the lattice responds to the TRS symmetry breaking due to AIAO order of Ir moments, and below 14~K to the ordering of Nd moments are discussed elsewhere~\cite{Xu2024}.

\subsection{Crystal field excitation of Nd$^{3+}$}
\label{CF}

\begin{figure}
  \centering
  \includegraphics[width=0.5\textwidth]{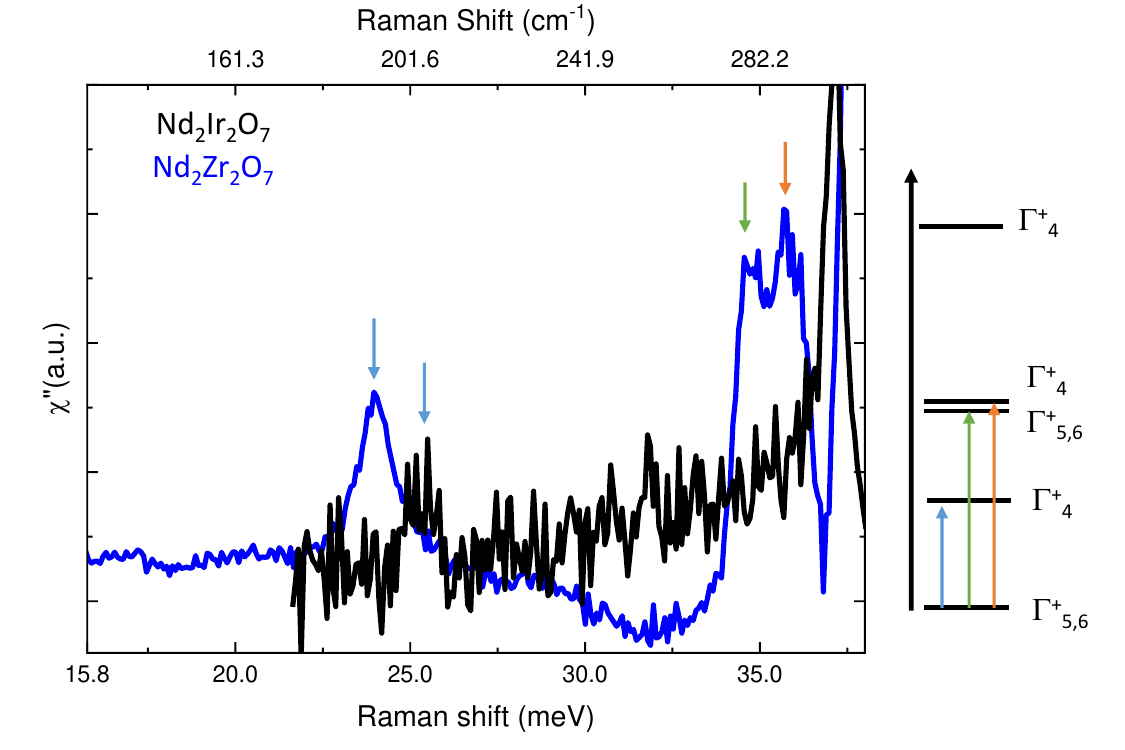}
  \caption{(a) Raman spectra of \nzo\ (blue line) and \nio\ (black line) in the region of the  two lowest frequency crystal electric field excitations (CEF). For \nzo\ we present  $\chi''$(CEF)=$\chi''$(14~K)-$\chi''$(300~K), where $\chi''$(T) is total Raman intensity measured at temperature T in $(x,y)$ channel. This subtraction reveals the doublet of CEF at 34 and 36 meV. For \nio\ $\chi''$(CEF)= $\chi''(T) - \chi''(263 K)$ averaged between T=50 and 90~K.  \label{NZO_CEF}}
\end{figure}

In addition to the excitations of the $\Gamma$-point phonons, in the measured frequency range we observe crystal field excitations of  Nd$^{3+}$. 
The local symmetry of the Nd$^{3+},  J$= 9/2, site in both structures is D$_{3d}$, which results in the splitting of the orbital into five Kramers doublets (see Fig.~\ref{NZO_CEF})~\cite{Watahiki2011,Xu2015}. For Nd$_2$Zr$_2$O$_7$, neutron scattering observed  transitions at 23.4  meV, 35 meV and 106.2~meV. The transition observed  at 35 meV was extra broad and thus assigned to the unresolved $\Gamma^+_4$, $\Gamma^+_{5,6}$ doublet~\cite{Xu2015}.

The transitions between the ground state and excited state doublets are expected to be observed in the Raman scattering spectra. The doublet states for D$_{3d}$ corresponding to E$_g$ point group symmetry, where E$_g  \rightarrow$ E$_g$ transitions will be observed in E$_g \otimes$ E$_g$ = A$_{1g}$+A$_{2g}$+E$_g$ scattering channels, meaning they will be visible in both $(x,x)$ and $(x,y)$ spectra for the studied pyrochlores.  Regarding the intensity of the crystal field excitations, our expectations are that it will be at least one order of magnitude lower than that of the oxygen phonons, as was already demonstrated for Pr$_2$Ir$_2$O$_7$ and Pr$_2$Zr$_2$O$_7$~\cite{Xu2021cef,Xu2022phonons}. One distinct property of CEF excitations in Raman scattering which allows us to distinguish them from other excitations is a strong temperature dependence of the scattering on temperature: As a result, these excitations are observed as sharp peaks only at temperatures lower than the energy of the excitation~\cite{Cardona2000}.

Due to the high overall intensity of Raman scattering in \nzo, the lower frequency CEF transitions can be relatively easily identified  using the comparison between high and low temperature spectra. A peak at 23.5 meV of a transition from the ground state to the first excited state $\Gamma^+_{5,6} \rightarrow \Gamma^+_4$ is well distinguished in the spectra below 160~K. CEF in the range of 35 meV appear as a double ``wing'' of the strong T$_{2g}^{(1)}$ phonon band clearly observed in the $(x,y)$ channel (Fig.~\ref{NZO_Raman}) on temperature lowering. Fig.~\ref{NZO_CEF} shows a result of the subtraction of $\chi''(T=14 K) - \chi''(300 K)$ in the $(x,y)$ channel for Nd$_2$Zr$_2$O$_7$. The doublet of excitations at 34.4~meV and 35.2~meV is revealed by the subtraction. 
These peaks correspond to the transitions from the ground state to the second and third excited states $\Gamma^+_{5,6}$ and  $\Gamma^+_{4}$,  which were expected by calculations, and observed by neutron scattering as a single broad band~\cite{Xu2015}.  

Since the overall intensity of the \nio\ spectra is two orders of magnitude lower than that of \nzo, detecting crystal field excitations becomes a challenge. Thus we can observe clearly only the excitation to the first excited state $\Gamma^+_{5,6} \rightarrow \Gamma^+_4$ found at at 25.3 meV (see Fig.~\ref{NZO_CEF}). This spectrum was obtained by averaging $\chi''(T) - \chi''(263 K)$ for the temperatures between 
50 and 90 K, using the fact that crystal field excitations show basically no temperature dependence of their frequencies.
The observed CEF is in agreement with neutron scattering measurements for   Nd$_2$Ir$_2$O$_7$, where transitions at 26 meV and 41 meV were observed, with the second peak being broader~\cite{Watahiki2011}. In our Raman scattering measurements the transitions to the second excited states couldn't be distinguished from the phonon scattering.

\section{Discussion}

The temperature dependence of the Raman-active $\Gamma$-point phonons in \nzo\ and \nio\ suggests that in both compounds the lattice is interacting with the other degrees of freedom and the scattering of phonons is determined by factors beyond phonon-phonon scattering as described by the Klemens model. The comparison highlights the dominance of a different scattering channel for the insulating \nzo\ versus semimetallic  \nio. 

In \nzo\ frequencies of the phonons harden on cooling down to approximately 100~K, and either stay temperature independent, or soften on cooling below this temperature, with the softening being the most pronounced for the T$^{(3)}_{2g}$ phonon. The observed hardening of the phonons down to 100~K is in agreement with the decrease of the lattice constant $a$ on cooling~\cite{Xu2015,Kim2012}.  The temperature dependence of the width of the phonons follows the shape determined by the phonon-phonon scattering as expected by the Klemens model~\cite{Kim2012}, but deviates from it around 100~K, suggesting an additional scattering channel in this relatively narrow temperature range. This characteristic temperature range around 100~K was also identified in \nzo\ heat capacity measurements~\cite{Xu2015,Hatnean2015} as a broad feature of a Schottky-like anomaly.  It was interpreted as a magnetic contribution associated with the depopulation of the first excited crystal field level at 23.5~meV. It is natural to assign the feature in the heat capacity and in the temperature dependence of the phonon scattering to the same effect, pointing to the scattering of phonons on crystal field excitations. The scattering of the crystal field excitations on the phonons is a known effect~\cite{Cardona2000}. To the best of our knowledge, the effect on the scattering of the phonons has never been detected yet, with the \nzo\ apparently providing the optimum system to detect this effect due to the convenient temperature scale~\cite{Cardona2000}. While the scattering of phonons on crystal field excitations is not discussed explicitly, this effect points to scattering channels of magnetic origin which can exist for phonons in rare-earth pyrochlores~\cite{Seth2022}


The large width of \nio\ phonons at room temperature compared to other pyrochlore iridates for which data are available~\cite{Ueda2019,Xu2022phonons,Rosalin_PIO_2023,Rosalin_AIO_2024} and the unusually fast decrease of scattering with the lowering of temperature for the two lower frequency phonons also suggests scattering  beyond the phonon-phonon one. Similarly to \pio, we suggest that electron-phonon scattering is the dominant process ~\cite{Xu2022phonons, Rosalin_PIO_2023}: A decrease of scattering occurs due to the relevant electronic levels depopulation on cooling. Both in \nio\ and \pio\ the phonons are only broad at high temperatures, with their scattering decreases below about 150~K.
 Interestingly, this temperature range around 150~K is close to the temperature range where other iridates go through the metal insulator transition~\cite{Witczak2014}.



Of a special interest is the comparison of our results on Nd-based materials with Pr-based ones.
Our previous Raman scattering studies of Pr-based compounds demonstrated a splitting of the E$_g$ phonon in \pzo\ and an extreme broadening of this phonon in \pio, which cannot be explained by vibronic coupling due to an energy overlap with a crystal field excitation~\cite{Xu2021cef,Xu2022phonons}. In contrast, E$_g$ phonons in Nd-based pyrochlores studied here do not show any splitting or broadening different from that of T$_{2g}$ phonons. A coupling of the E$_g$ lattice degree of freedom with magnetic degree of freedom in non-Kramers pyrochlores,  which does not require an overlap of energies of magnetic excitations and phonons, suggested in Ref.~\cite{Seth2022}, can provide an explanation for our observations. The respective splitting of the ground state crystal field level in \pzo\ has been detected by neutron scattering~\cite{Wen2017} and Raman scattering spectroscopy~\cite{Xu2021cef} at low temperatures. \pio\ can posses similar properties, however the overlap in the spectral range with electronic scattering prevents the detection of the relevant crystal field transitions~\cite{Nikolic2024}.  
The detection of the coupling proposed in Ref.~\cite{Seth2022} in the high-temperature paramagnetic regime of the rare earth pyrochlores can be the first step to the identification the scattering on the excitations in quantum spin ice state of these materials.   

While the phonon behavior of these two materials shows unexpected behavior, crystal field levels follow the calculated energies~\cite{Xu2015}. For \nzo\ Raman scattering measurements allow us to resolve the expected from the calculation doublet of transitions to the second and third excited states $\Gamma^+_{5,6}$ and  $\Gamma^+_{4}$ at  34.4~meV and 35.2~meV. For \nio, due to very low intensities of the spectra in a semimetal and an overlap with the phonons, we can only identify the position of the first excited state. The width of the excitation is similar to that of \nzo, suggesting that the lifetime of these localized excitations are not very affected by the difference in the electronic structure of the materials

\section{Conclusions}

In this work we present the evolution of phonon and crystal field Raman scattering in \nzo\ and \nio\ on lowering temperature in the temperature range where both materials are parmagnetic. We demonstrate that in \nzo\ the lattice reacts to the depopulation of the lowest crystal field level, providing an additional scattering channel for phonons in the relevant temperature range, in addition to the phonon-phonon scattering described by Klemens model. A correlation between thermal contraction and CEF levels depopulation also points to the coupling with between lattice and crystal field degrees of freedom in \nzo.
Raman scattering spectra allow us to measure the positions of CEF in \nzo\ with high precision and resolve a doublet which was expected from the calculations. 

Two lower frequency \nio\ phonons (T$_{2g}^{(1)}$ and E$_g$) experience dramatic decreases of scattering on cooling from room temperature down to T$_N^{Nd}$, interpreted as  a consequence of electron-phonon scattering and the decrease of the thermal population of relevant electronic levels.

\begin{acknowledgments}
The authors are thankful to Y. Yang  for useful discussions. This work at the Institute for Quantum Matter was funded by the U.S. Department of Energy, Office of Science, Basic Energy Sciences under Awards No.~DE-SC0019331 and DE-SC0024469. This work was partially supported by JST-MIRAI Program (JPMJMI20A1), JST-ASPIRE Program (JPMJAP2317) and by the fund made by Canadian Institute for Advanced Research.
SDW and EZ acknowledge support from the US Department of Energy (DOE), Office of Basic Energy Sciences, Division
of Materials Sciences and Engineering under Grant No. DE-SC0017752.  SDW and EZ's work used facilities supported via the UC Santa Barbara NSF Quantum Foundry funded via the Q-AMASE-i program under award DMR-1906325.
\end{acknowledgments}

\bibliography{PyrochloresNIO}

\begin{thebibliography}{38}%
\makeatletter
\providecommand \@ifxundefined [1]{%
 \@ifx{#1\undefined}
}%
\providecommand \@ifnum [1]{%
 \ifnum #1\expandafter \@firstoftwo
 \else \expandafter \@secondoftwo
 \fi
}%
\providecommand \@ifx [1]{%
 \ifx #1\expandafter \@firstoftwo
 \else \expandafter \@secondoftwo
 \fi
}%
\providecommand \natexlab [1]{#1}%
\providecommand \enquote  [1]{``#1''}%
\providecommand \bibnamefont  [1]{#1}%
\providecommand \bibfnamefont [1]{#1}%
\providecommand \citenamefont [1]{#1}%
\providecommand \href@noop [0]{\@secondoftwo}%
\providecommand \href [0]{\begingroup \@sanitize@url \@href}%
\providecommand \@href[1]{\@@startlink{#1}\@@href}%
\providecommand \@@href[1]{\endgroup#1\@@endlink}%
\providecommand \@sanitize@url [0]{\catcode `\\12\catcode `\$12\catcode `\&12\catcode `\#12\catcode `\^12\catcode `\_12\catcode `\%12\relax}%
\providecommand \@@startlink[1]{}%
\providecommand \@@endlink[0]{}%
\providecommand \url  [0]{\begingroup\@sanitize@url \@url }%
\providecommand \@url [1]{\endgroup\@href {#1}{\urlprefix }}%
\providecommand \urlprefix  [0]{URL }%
\providecommand \Eprint [0]{\href }%
\providecommand \doibase [0]{https://doi.org/}%
\providecommand \selectlanguage [0]{\@gobble}%
\providecommand \bibinfo  [0]{\@secondoftwo}%
\providecommand \bibfield  [0]{\@secondoftwo}%
\providecommand \translation [1]{[#1]}%
\providecommand \BibitemOpen [0]{}%
\providecommand \bibitemStop [0]{}%
\providecommand \bibitemNoStop [0]{.\EOS\space}%
\providecommand \EOS [0]{\spacefactor3000\relax}%
\providecommand \BibitemShut  [1]{\csname bibitem#1\endcsname}%
\let\auto@bib@innerbib\@empty
\bibitem [{\citenamefont {Gardner}\ \emph {et~al.}(2010)\citenamefont {Gardner}, \citenamefont {Gingras},\ and\ \citenamefont {Greedan}}]{Gardner2010}%
  \BibitemOpen
  \bibfield  {author} {\bibinfo {author} {\bibfnamefont {J.~S.}\ \bibnamefont {Gardner}}, \bibinfo {author} {\bibfnamefont {M.~J.~P.}\ \bibnamefont {Gingras}},\ and\ \bibinfo {author} {\bibfnamefont {J.~E.}\ \bibnamefont {Greedan}},\ }\href {https://doi.org/10.1103/RevModPhys.82.53} {\bibfield  {journal} {\bibinfo  {journal} {Rev. Mod. Phys.}\ }\textbf {\bibinfo {volume} {82}},\ \bibinfo {pages} {53} (\bibinfo {year} {2010})}\BibitemShut {NoStop}%
\bibitem [{\citenamefont {Gingras}\ and\ \citenamefont {McClarty}(2014)}]{Gingras2014}%
  \BibitemOpen
  \bibfield  {author} {\bibinfo {author} {\bibfnamefont {M.~J.}\ \bibnamefont {Gingras}}\ and\ \bibinfo {author} {\bibfnamefont {P.~A.}\ \bibnamefont {McClarty}},\ }\href {https://doi.org/10.1088/0034-4885/77/5/056501} {\bibfield  {journal} {\bibinfo  {journal} {Reports on Progress in Physics}\ }\textbf {\bibinfo {volume} {77}},\ \bibinfo {pages} {056501} (\bibinfo {year} {2014})}\BibitemShut {NoStop}%
\bibitem [{\citenamefont {Rau}\ and\ \citenamefont {Gingras}(2019)}]{Rau2019}%
  \BibitemOpen
  \bibfield  {author} {\bibinfo {author} {\bibfnamefont {J.~G.}\ \bibnamefont {Rau}}\ and\ \bibinfo {author} {\bibfnamefont {M.~J.}\ \bibnamefont {Gingras}},\ }\href {https://doi.org/10.1146/annurev-conmatphys-022317-110520} {\bibfield  {journal} {\bibinfo  {journal} {Annual Review of Condensed Matter Physics}\ }\textbf {\bibinfo {volume} {10}},\ \bibinfo {pages} {357} (\bibinfo {year} {2019})}\BibitemShut {NoStop}%
\bibitem [{\citenamefont {Gaudet}\ \emph {et~al.}(2019)\citenamefont {Gaudet}, \citenamefont {Smith}, \citenamefont {Dudemaine}, \citenamefont {Beare}, \citenamefont {Buhariwalla}, \citenamefont {Butch}, \citenamefont {Stone}, \citenamefont {Kolesnikov}, \citenamefont {Xu}, \citenamefont {Yahne}, \citenamefont {Ross}, \citenamefont {Marjerrison}, \citenamefont {Garrett}, \citenamefont {Luke}, \citenamefont {Bianchi},\ and\ \citenamefont {Gaulin}}]{Gaudet_CZO_2019}%
  \BibitemOpen
  \bibfield  {author} {\bibinfo {author} {\bibfnamefont {J.}~\bibnamefont {Gaudet}}, \bibinfo {author} {\bibfnamefont {E.~M.}\ \bibnamefont {Smith}}, \bibinfo {author} {\bibfnamefont {J.}~\bibnamefont {Dudemaine}}, \bibinfo {author} {\bibfnamefont {J.}~\bibnamefont {Beare}}, \bibinfo {author} {\bibfnamefont {C.~R.~C.}\ \bibnamefont {Buhariwalla}}, \bibinfo {author} {\bibfnamefont {N.~P.}\ \bibnamefont {Butch}}, \bibinfo {author} {\bibfnamefont {M.~B.}\ \bibnamefont {Stone}}, \bibinfo {author} {\bibfnamefont {A.~I.}\ \bibnamefont {Kolesnikov}}, \bibinfo {author} {\bibfnamefont {G.}~\bibnamefont {Xu}}, \bibinfo {author} {\bibfnamefont {D.~R.}\ \bibnamefont {Yahne}}, \bibinfo {author} {\bibfnamefont {K.~A.}\ \bibnamefont {Ross}}, \bibinfo {author} {\bibfnamefont {C.~A.}\ \bibnamefont {Marjerrison}}, \bibinfo {author} {\bibfnamefont {J.~D.}\ \bibnamefont {Garrett}}, \bibinfo {author} {\bibfnamefont {G.~M.}\ \bibnamefont {Luke}}, \bibinfo {author} {\bibfnamefont {A.~D.}\ \bibnamefont {Bianchi}},\ and\ \bibinfo
  {author} {\bibfnamefont {B.~D.}\ \bibnamefont {Gaulin}},\ }\href {https://doi.org/10.1103/PhysRevLett.122.187201} {\bibfield  {journal} {\bibinfo  {journal} {Phys. Rev. Lett.}\ }\textbf {\bibinfo {volume} {122}},\ \bibinfo {pages} {187201} (\bibinfo {year} {2019})}\BibitemShut {NoStop}%
\bibitem [{\citenamefont {Gao}\ \emph {et~al.}(2019)\citenamefont {Gao}, \citenamefont {Chen}, \citenamefont {Tam}, \citenamefont {Huang}, \citenamefont {Sasmal}, \citenamefont {Adroja}, \citenamefont {Ye}, \citenamefont {Cao}, \citenamefont {Sala}, \citenamefont {Stone} \emph {et~al.}}]{Gao_CZO_2019}%
  \BibitemOpen
  \bibfield  {author} {\bibinfo {author} {\bibfnamefont {B.}~\bibnamefont {Gao}}, \bibinfo {author} {\bibfnamefont {T.}~\bibnamefont {Chen}}, \bibinfo {author} {\bibfnamefont {D.~W.}\ \bibnamefont {Tam}}, \bibinfo {author} {\bibfnamefont {C.-L.}\ \bibnamefont {Huang}}, \bibinfo {author} {\bibfnamefont {K.}~\bibnamefont {Sasmal}}, \bibinfo {author} {\bibfnamefont {D.~T.}\ \bibnamefont {Adroja}}, \bibinfo {author} {\bibfnamefont {F.}~\bibnamefont {Ye}}, \bibinfo {author} {\bibfnamefont {H.}~\bibnamefont {Cao}}, \bibinfo {author} {\bibfnamefont {G.}~\bibnamefont {Sala}}, \bibinfo {author} {\bibfnamefont {M.~B.}\ \bibnamefont {Stone}}, \emph {et~al.},\ }\href@noop {} {\bibfield  {journal} {\bibinfo  {journal} {Nature Physics}\ }\textbf {\bibinfo {volume} {15}},\ \bibinfo {pages} {1052} (\bibinfo {year} {2019})}\BibitemShut {NoStop}%
\bibitem [{\citenamefont {Smith}\ \emph {et~al.}(2022)\citenamefont {Smith}, \citenamefont {Benton}, \citenamefont {Yahne}, \citenamefont {Placke}, \citenamefont {Sch\"afer}, \citenamefont {Gaudet}, \citenamefont {Dudemaine}, \citenamefont {Fitterman}, \citenamefont {Beare}, \citenamefont {Wildes}, \citenamefont {Bhattacharya}, \citenamefont {DeLazzer}, \citenamefont {Buhariwalla}, \citenamefont {Butch}, \citenamefont {Movshovich}, \citenamefont {Garrett}, \citenamefont {Marjerrison}, \citenamefont {Clancy}, \citenamefont {Kermarrec}, \citenamefont {Luke}, \citenamefont {Bianchi}, \citenamefont {Ross},\ and\ \citenamefont {Gaulin}}]{Smith_CZO_2022}%
  \BibitemOpen
  \bibfield  {author} {\bibinfo {author} {\bibfnamefont {E.~M.}\ \bibnamefont {Smith}}, \bibinfo {author} {\bibfnamefont {O.}~\bibnamefont {Benton}}, \bibinfo {author} {\bibfnamefont {D.~R.}\ \bibnamefont {Yahne}}, \bibinfo {author} {\bibfnamefont {B.}~\bibnamefont {Placke}}, \bibinfo {author} {\bibfnamefont {R.}~\bibnamefont {Sch\"afer}}, \bibinfo {author} {\bibfnamefont {J.}~\bibnamefont {Gaudet}}, \bibinfo {author} {\bibfnamefont {J.}~\bibnamefont {Dudemaine}}, \bibinfo {author} {\bibfnamefont {A.}~\bibnamefont {Fitterman}}, \bibinfo {author} {\bibfnamefont {J.}~\bibnamefont {Beare}}, \bibinfo {author} {\bibfnamefont {A.~R.}\ \bibnamefont {Wildes}}, \bibinfo {author} {\bibfnamefont {S.}~\bibnamefont {Bhattacharya}}, \bibinfo {author} {\bibfnamefont {T.}~\bibnamefont {DeLazzer}}, \bibinfo {author} {\bibfnamefont {C.~R.~C.}\ \bibnamefont {Buhariwalla}}, \bibinfo {author} {\bibfnamefont {N.~P.}\ \bibnamefont {Butch}}, \bibinfo {author} {\bibfnamefont {R.}~\bibnamefont {Movshovich}}, \bibinfo {author}
  {\bibfnamefont {J.~D.}\ \bibnamefont {Garrett}}, \bibinfo {author} {\bibfnamefont {C.~A.}\ \bibnamefont {Marjerrison}}, \bibinfo {author} {\bibfnamefont {J.~P.}\ \bibnamefont {Clancy}}, \bibinfo {author} {\bibfnamefont {E.}~\bibnamefont {Kermarrec}}, \bibinfo {author} {\bibfnamefont {G.~M.}\ \bibnamefont {Luke}}, \bibinfo {author} {\bibfnamefont {A.~D.}\ \bibnamefont {Bianchi}}, \bibinfo {author} {\bibfnamefont {K.~A.}\ \bibnamefont {Ross}},\ and\ \bibinfo {author} {\bibfnamefont {B.~D.}\ \bibnamefont {Gaulin}},\ }\href {https://doi.org/10.1103/PhysRevX.12.021015} {\bibfield  {journal} {\bibinfo  {journal} {Phys. Rev. X}\ }\textbf {\bibinfo {volume} {12}},\ \bibinfo {pages} {021015} (\bibinfo {year} {2022})}\BibitemShut {NoStop}%
\bibitem [{\citenamefont {Bhardwaj}\ \emph {et~al.}(2022)\citenamefont {Bhardwaj}, \citenamefont {Zhang}, \citenamefont {Yan}, \citenamefont {Moessner}, \citenamefont {Nevidomskyy},\ and\ \citenamefont {Changlani}}]{Bhardwaj_CZO_2022}%
  \BibitemOpen
  \bibfield  {author} {\bibinfo {author} {\bibfnamefont {A.}~\bibnamefont {Bhardwaj}}, \bibinfo {author} {\bibfnamefont {S.}~\bibnamefont {Zhang}}, \bibinfo {author} {\bibfnamefont {H.}~\bibnamefont {Yan}}, \bibinfo {author} {\bibfnamefont {R.}~\bibnamefont {Moessner}}, \bibinfo {author} {\bibfnamefont {A.~H.}\ \bibnamefont {Nevidomskyy}},\ and\ \bibinfo {author} {\bibfnamefont {H.~J.}\ \bibnamefont {Changlani}},\ }\href@noop {} {\bibfield  {journal} {\bibinfo  {journal} {npj Quantum Materials}\ }\textbf {\bibinfo {volume} {7}},\ \bibinfo {pages} {51} (\bibinfo {year} {2022})}\BibitemShut {NoStop}%
\bibitem [{\citenamefont {L\'eger}\ \emph {et~al.}(2021)\citenamefont {L\'eger}, \citenamefont {Lhotel}, \citenamefont {Ciomaga~Hatnean}, \citenamefont {Ollivier}, \citenamefont {Wildes}, \citenamefont {Raymond}, \citenamefont {Ressouche}, \citenamefont {Balakrishnan},\ and\ \citenamefont {Petit}}]{Lhotel2021}%
  \BibitemOpen
  \bibfield  {author} {\bibinfo {author} {\bibfnamefont {M.}~\bibnamefont {L\'eger}}, \bibinfo {author} {\bibfnamefont {E.}~\bibnamefont {Lhotel}}, \bibinfo {author} {\bibfnamefont {M.}~\bibnamefont {Ciomaga~Hatnean}}, \bibinfo {author} {\bibfnamefont {J.}~\bibnamefont {Ollivier}}, \bibinfo {author} {\bibfnamefont {A.~R.}\ \bibnamefont {Wildes}}, \bibinfo {author} {\bibfnamefont {S.}~\bibnamefont {Raymond}}, \bibinfo {author} {\bibfnamefont {E.}~\bibnamefont {Ressouche}}, \bibinfo {author} {\bibfnamefont {G.}~\bibnamefont {Balakrishnan}},\ and\ \bibinfo {author} {\bibfnamefont {S.}~\bibnamefont {Petit}},\ }\href {https://doi.org/10.1103/PhysRevLett.126.247201} {\bibfield  {journal} {\bibinfo  {journal} {Phys. Rev. Lett.}\ }\textbf {\bibinfo {volume} {126}},\ \bibinfo {pages} {247201} (\bibinfo {year} {2021})}\BibitemShut {NoStop}%
\bibitem [{\citenamefont {Xu}\ \emph {et~al.}(2020)\citenamefont {Xu}, \citenamefont {Benton}, \citenamefont {Islam}, \citenamefont {Guidi}, \citenamefont {Ehlers},\ and\ \citenamefont {Lake}}]{Xu_NZO_2020}%
  \BibitemOpen
  \bibfield  {author} {\bibinfo {author} {\bibfnamefont {J.}~\bibnamefont {Xu}}, \bibinfo {author} {\bibfnamefont {O.}~\bibnamefont {Benton}}, \bibinfo {author} {\bibfnamefont {A.}~\bibnamefont {Islam}}, \bibinfo {author} {\bibfnamefont {T.}~\bibnamefont {Guidi}}, \bibinfo {author} {\bibfnamefont {G.}~\bibnamefont {Ehlers}},\ and\ \bibinfo {author} {\bibfnamefont {B.}~\bibnamefont {Lake}},\ }\href@noop {} {\bibfield  {journal} {\bibinfo  {journal} {Physical Review Letters}\ }\textbf {\bibinfo {volume} {124}},\ \bibinfo {pages} {097203} (\bibinfo {year} {2020})}\BibitemShut {NoStop}%
\bibitem [{\citenamefont {Witczak-Krempa}\ and\ \citenamefont {Kim}(2012)}]{Witczak2012}%
  \BibitemOpen
  \bibfield  {author} {\bibinfo {author} {\bibfnamefont {W.}~\bibnamefont {Witczak-Krempa}}\ and\ \bibinfo {author} {\bibfnamefont {Y.~B.}\ \bibnamefont {Kim}},\ }\href {https://doi.org/10.1103/PhysRevB.85.045124} {\bibfield  {journal} {\bibinfo  {journal} {Phys. Rev. B}\ }\textbf {\bibinfo {volume} {85}},\ \bibinfo {pages} {045124} (\bibinfo {year} {2012})}\BibitemShut {NoStop}%
\bibitem [{\citenamefont {Matsuhira}\ \emph {et~al.}(2011)\citenamefont {Matsuhira}, \citenamefont {Wakeshima}, \citenamefont {Hinatsu},\ and\ \citenamefont {Takagi}}]{Matsuhira2011}%
  \BibitemOpen
  \bibfield  {author} {\bibinfo {author} {\bibfnamefont {K.}~\bibnamefont {Matsuhira}}, \bibinfo {author} {\bibfnamefont {M.}~\bibnamefont {Wakeshima}}, \bibinfo {author} {\bibfnamefont {Y.}~\bibnamefont {Hinatsu}},\ and\ \bibinfo {author} {\bibfnamefont {S.}~\bibnamefont {Takagi}},\ }\href {https://doi.org/10.1143/jpsj.80.094701} {\bibfield  {journal} {\bibinfo  {journal} {Journal of the Physical Society of Japan}\ }\textbf {\bibinfo {volume} {80}},\ \bibinfo {pages} {094701} (\bibinfo {year} {2011})}\BibitemShut {NoStop}%
\bibitem [{\citenamefont {Ueda}\ \emph {et~al.}(2012)\citenamefont {Ueda}, \citenamefont {Fujioka}, \citenamefont {Takahashi}, \citenamefont {Suzuki}, \citenamefont {Ishiwata}, \citenamefont {Taguchi},\ and\ \citenamefont {Tokura}}]{Ueda2012}%
  \BibitemOpen
  \bibfield  {author} {\bibinfo {author} {\bibfnamefont {K.}~\bibnamefont {Ueda}}, \bibinfo {author} {\bibfnamefont {J.}~\bibnamefont {Fujioka}}, \bibinfo {author} {\bibfnamefont {Y.}~\bibnamefont {Takahashi}}, \bibinfo {author} {\bibfnamefont {T.}~\bibnamefont {Suzuki}}, \bibinfo {author} {\bibfnamefont {S.}~\bibnamefont {Ishiwata}}, \bibinfo {author} {\bibfnamefont {Y.}~\bibnamefont {Taguchi}},\ and\ \bibinfo {author} {\bibfnamefont {Y.}~\bibnamefont {Tokura}},\ }\href {https://doi.org/10.1103/PhysRevLett.109.136402} {\bibfield  {journal} {\bibinfo  {journal} {Phys. Rev. Lett.}\ }\textbf {\bibinfo {volume} {109}},\ \bibinfo {pages} {136402} (\bibinfo {year} {2012})}\BibitemShut {NoStop}%
\bibitem [{\citenamefont {Wan}\ \emph {et~al.}(2011)\citenamefont {Wan}, \citenamefont {Turner}, \citenamefont {Vishwanath},\ and\ \citenamefont {Savrasov}}]{Wan2011}%
  \BibitemOpen
  \bibfield  {author} {\bibinfo {author} {\bibfnamefont {X.}~\bibnamefont {Wan}}, \bibinfo {author} {\bibfnamefont {A.~M.}\ \bibnamefont {Turner}}, \bibinfo {author} {\bibfnamefont {A.}~\bibnamefont {Vishwanath}},\ and\ \bibinfo {author} {\bibfnamefont {S.~Y.}\ \bibnamefont {Savrasov}},\ }\href {https://doi.org/10.1103/PhysRevB.83.205101} {\bibfield  {journal} {\bibinfo  {journal} {Physical Review B}\ }\textbf {\bibinfo {volume} {83}},\ \bibinfo {pages} {205101} (\bibinfo {year} {2011})}\BibitemShut {NoStop}%
\bibitem [{\citenamefont {Witczak-Krempa}\ \emph {et~al.}(2014)\citenamefont {Witczak-Krempa}, \citenamefont {Chen}, \citenamefont {Kim},\ and\ \citenamefont {Balents}}]{Witczak2014}%
  \BibitemOpen
  \bibfield  {author} {\bibinfo {author} {\bibfnamefont {W.}~\bibnamefont {Witczak-Krempa}}, \bibinfo {author} {\bibfnamefont {G.}~\bibnamefont {Chen}}, \bibinfo {author} {\bibfnamefont {Y.~B.}\ \bibnamefont {Kim}},\ and\ \bibinfo {author} {\bibfnamefont {L.}~\bibnamefont {Balents}},\ }\href {https://doi.org/10.1146/annurev-conmatphys-020911-125138} {\bibfield  {journal} {\bibinfo  {journal} {Annual Review of Condensed Matter Physics}\ }\textbf {\bibinfo {volume} {5}},\ \bibinfo {pages} {57} (\bibinfo {year} {2014})}\BibitemShut {NoStop}%
\bibitem [{\citenamefont {Savary}\ and\ \citenamefont {Balents}(2016)}]{Savary2016}%
  \BibitemOpen
  \bibfield  {author} {\bibinfo {author} {\bibfnamefont {L.}~\bibnamefont {Savary}}\ and\ \bibinfo {author} {\bibfnamefont {L.}~\bibnamefont {Balents}},\ }\href@noop {} {\bibfield  {journal} {\bibinfo  {journal} {Reports on Progress in Physics}\ }\textbf {\bibinfo {volume} {80}},\ \bibinfo {pages} {016502} (\bibinfo {year} {2016})}\BibitemShut {NoStop}%
\bibitem [{\citenamefont {Nikoli\ifmmode~\acute{c}\else \'{c}\fi{}}\ \emph {et~al.}(2024)\citenamefont {Nikoli\ifmmode~\acute{c}\else \'{c}\fi{}}, \citenamefont {Xu}, \citenamefont {Ohtsuki}, \citenamefont {Elbert}, \citenamefont {Nakatsuji},\ and\ \citenamefont {Drichko}}]{Nikolic2024}%
  \BibitemOpen
  \bibfield  {author} {\bibinfo {author} {\bibfnamefont {P.}~\bibnamefont {Nikoli\ifmmode~\acute{c}\else \'{c}\fi{}}}, \bibinfo {author} {\bibfnamefont {Y.}~\bibnamefont {Xu}}, \bibinfo {author} {\bibfnamefont {T.}~\bibnamefont {Ohtsuki}}, \bibinfo {author} {\bibfnamefont {D.~C.}\ \bibnamefont {Elbert}}, \bibinfo {author} {\bibfnamefont {S.}~\bibnamefont {Nakatsuji}},\ and\ \bibinfo {author} {\bibfnamefont {N.}~\bibnamefont {Drichko}},\ }\href {https://doi.org/10.1103/PhysRevB.110.035148} {\bibfield  {journal} {\bibinfo  {journal} {Phys. Rev. B}\ }\textbf {\bibinfo {volume} {110}},\ \bibinfo {pages} {035148} (\bibinfo {year} {2024})}\BibitemShut {NoStop}%
\bibitem [{\citenamefont {Pearce}\ \emph {et~al.}(2022)\citenamefont {Pearce}, \citenamefont {G{\"o}tze}, \citenamefont {Szab{\'o}}, \citenamefont {Sikkenk}, \citenamefont {Lees}, \citenamefont {Boothroyd}, \citenamefont {Prabhakaran}, \citenamefont {Castelnovo},\ and\ \citenamefont {Goddard}}]{Pearce2022}%
  \BibitemOpen
  \bibfield  {author} {\bibinfo {author} {\bibfnamefont {M.~J.}\ \bibnamefont {Pearce}}, \bibinfo {author} {\bibfnamefont {K.}~\bibnamefont {G{\"o}tze}}, \bibinfo {author} {\bibfnamefont {A.}~\bibnamefont {Szab{\'o}}}, \bibinfo {author} {\bibfnamefont {T.}~\bibnamefont {Sikkenk}}, \bibinfo {author} {\bibfnamefont {M.~R.}\ \bibnamefont {Lees}}, \bibinfo {author} {\bibfnamefont {A.}~\bibnamefont {Boothroyd}}, \bibinfo {author} {\bibfnamefont {D.}~\bibnamefont {Prabhakaran}}, \bibinfo {author} {\bibfnamefont {C.}~\bibnamefont {Castelnovo}},\ and\ \bibinfo {author} {\bibfnamefont {P.}~\bibnamefont {Goddard}},\ }\href@noop {} {\bibfield  {journal} {\bibinfo  {journal} {Nature Communications}\ }\textbf {\bibinfo {volume} {13}},\ \bibinfo {pages} {444} (\bibinfo {year} {2022})}\BibitemShut {NoStop}%
\bibitem [{\citenamefont {Xu}\ \emph {et~al.}(2023)\citenamefont {Xu}, \citenamefont {Yang}, \citenamefont {Teyssier}, \citenamefont {Ohtsuki}, \citenamefont {Qiu}, \citenamefont {Nakatsuji}, \citenamefont {van~der Marel}, \citenamefont {Perkins},\ and\ \citenamefont {Drichko}}]{Xu2024}%
  \BibitemOpen
  \bibfield  {author} {\bibinfo {author} {\bibfnamefont {Y.}~\bibnamefont {Xu}}, \bibinfo {author} {\bibfnamefont {Y.}~\bibnamefont {Yang}}, \bibinfo {author} {\bibfnamefont {J.}~\bibnamefont {Teyssier}}, \bibinfo {author} {\bibfnamefont {T.}~\bibnamefont {Ohtsuki}}, \bibinfo {author} {\bibfnamefont {Y.}~\bibnamefont {Qiu}}, \bibinfo {author} {\bibfnamefont {S.}~\bibnamefont {Nakatsuji}}, \bibinfo {author} {\bibfnamefont {D.}~\bibnamefont {van~der Marel}}, \bibinfo {author} {\bibfnamefont {N.~B.}\ \bibnamefont {Perkins}},\ and\ \bibinfo {author} {\bibfnamefont {N.}~\bibnamefont {Drichko}},\ }\href@noop {} {\bibfield  {journal} {\bibinfo  {journal} {arXiv preprint arXiv:2302.00579}\ } (\bibinfo {year} {2023})}\BibitemShut {NoStop}%
\bibitem [{\citenamefont {Tang}\ \emph {et~al.}(2023)\citenamefont {Tang}, \citenamefont {Gritsenko}, \citenamefont {Kimura}, \citenamefont {Bhattacharjee}, \citenamefont {Sakai}, \citenamefont {Fu}, \citenamefont {Takeda}, \citenamefont {Man}, \citenamefont {Sugawara}, \citenamefont {Matsumoto} \emph {et~al.}}]{Tang2023}%
  \BibitemOpen
  \bibfield  {author} {\bibinfo {author} {\bibfnamefont {N.}~\bibnamefont {Tang}}, \bibinfo {author} {\bibfnamefont {Y.}~\bibnamefont {Gritsenko}}, \bibinfo {author} {\bibfnamefont {K.}~\bibnamefont {Kimura}}, \bibinfo {author} {\bibfnamefont {S.}~\bibnamefont {Bhattacharjee}}, \bibinfo {author} {\bibfnamefont {A.}~\bibnamefont {Sakai}}, \bibinfo {author} {\bibfnamefont {M.}~\bibnamefont {Fu}}, \bibinfo {author} {\bibfnamefont {H.}~\bibnamefont {Takeda}}, \bibinfo {author} {\bibfnamefont {H.}~\bibnamefont {Man}}, \bibinfo {author} {\bibfnamefont {K.}~\bibnamefont {Sugawara}}, \bibinfo {author} {\bibfnamefont {Y.}~\bibnamefont {Matsumoto}}, \emph {et~al.},\ }\href@noop {} {\bibfield  {journal} {\bibinfo  {journal} {Nature Physics}\ }\textbf {\bibinfo {volume} {19}},\ \bibinfo {pages} {92} (\bibinfo {year} {2023})}\BibitemShut {NoStop}%
\bibitem [{\citenamefont {Xu}\ \emph {et~al.}(2021)\citenamefont {Xu}, \citenamefont {Man}, \citenamefont {Tang}, \citenamefont {Baidya}, \citenamefont {Zhang}, \citenamefont {Nakatsuji}, \citenamefont {Vanderbilt},\ and\ \citenamefont {Drichko}}]{Xu2021cef}%
  \BibitemOpen
  \bibfield  {author} {\bibinfo {author} {\bibfnamefont {Y.}~\bibnamefont {Xu}}, \bibinfo {author} {\bibfnamefont {H.}~\bibnamefont {Man}}, \bibinfo {author} {\bibfnamefont {N.}~\bibnamefont {Tang}}, \bibinfo {author} {\bibfnamefont {S.}~\bibnamefont {Baidya}}, \bibinfo {author} {\bibfnamefont {H.}~\bibnamefont {Zhang}}, \bibinfo {author} {\bibfnamefont {S.}~\bibnamefont {Nakatsuji}}, \bibinfo {author} {\bibfnamefont {D.}~\bibnamefont {Vanderbilt}},\ and\ \bibinfo {author} {\bibfnamefont {N.}~\bibnamefont {Drichko}},\ }\href {https://doi.org/10.1103/PhysRevB.104.075125} {\bibfield  {journal} {\bibinfo  {journal} {Phys. Rev. B}\ }\textbf {\bibinfo {volume} {104}},\ \bibinfo {pages} {075125} (\bibinfo {year} {2021})}\BibitemShut {NoStop}%
\bibitem [{\citenamefont {Gaudet}\ \emph {et~al.}(2018)\citenamefont {Gaudet}, \citenamefont {Hallas}, \citenamefont {Buhariwalla}, \citenamefont {Sala}, \citenamefont {Stone}, \citenamefont {Tachibana}, \citenamefont {Baroudi}, \citenamefont {Cava},\ and\ \citenamefont {Gaulin}}]{Gaudet2018}%
  \BibitemOpen
  \bibfield  {author} {\bibinfo {author} {\bibfnamefont {J.}~\bibnamefont {Gaudet}}, \bibinfo {author} {\bibfnamefont {A.~M.}\ \bibnamefont {Hallas}}, \bibinfo {author} {\bibfnamefont {C.~R.~C.}\ \bibnamefont {Buhariwalla}}, \bibinfo {author} {\bibfnamefont {G.}~\bibnamefont {Sala}}, \bibinfo {author} {\bibfnamefont {M.~B.}\ \bibnamefont {Stone}}, \bibinfo {author} {\bibfnamefont {M.}~\bibnamefont {Tachibana}}, \bibinfo {author} {\bibfnamefont {K.}~\bibnamefont {Baroudi}}, \bibinfo {author} {\bibfnamefont {R.~J.}\ \bibnamefont {Cava}},\ and\ \bibinfo {author} {\bibfnamefont {B.~D.}\ \bibnamefont {Gaulin}},\ }\href {https://doi.org/10.1103/PhysRevB.98.014419} {\bibfield  {journal} {\bibinfo  {journal} {Phys. Rev. B}\ }\textbf {\bibinfo {volume} {98}},\ \bibinfo {pages} {014419} (\bibinfo {year} {2018})}\BibitemShut {NoStop}%
\bibitem [{\citenamefont {Thalmeier}\ and\ \citenamefont {Fulde}(1982)}]{Thalmeier1982}%
  \BibitemOpen
  \bibfield  {author} {\bibinfo {author} {\bibfnamefont {P.}~\bibnamefont {Thalmeier}}\ and\ \bibinfo {author} {\bibfnamefont {P.}~\bibnamefont {Fulde}},\ }\href {https://doi.org/10.1103/physrevlett.49.1588} {\bibfield  {journal} {\bibinfo  {journal} {Physical Review Letters}\ }\textbf {\bibinfo {volume} {49}},\ \bibinfo {pages} {1588} (\bibinfo {year} {1982})}\BibitemShut {NoStop}%
\bibitem [{\citenamefont {Sohn}\ \emph {et~al.}(2017)\citenamefont {Sohn}, \citenamefont {Kim}, \citenamefont {Sandilands}, \citenamefont {Hien}, \citenamefont {Kim}, \citenamefont {Park}, \citenamefont {Kim}, \citenamefont {Moon}, \citenamefont {Yamaura}, \citenamefont {Hiroi} \emph {et~al.}}]{Sohn_COO_2017}%
  \BibitemOpen
  \bibfield  {author} {\bibinfo {author} {\bibfnamefont {C.~H.}\ \bibnamefont {Sohn}}, \bibinfo {author} {\bibfnamefont {C.~H.}\ \bibnamefont {Kim}}, \bibinfo {author} {\bibfnamefont {L.~J.}\ \bibnamefont {Sandilands}}, \bibinfo {author} {\bibfnamefont {N.~T.~M.}\ \bibnamefont {Hien}}, \bibinfo {author} {\bibfnamefont {S.~Y.}\ \bibnamefont {Kim}}, \bibinfo {author} {\bibfnamefont {H.~J.}\ \bibnamefont {Park}}, \bibinfo {author} {\bibfnamefont {K.~W.}\ \bibnamefont {Kim}}, \bibinfo {author} {\bibfnamefont {S.}~\bibnamefont {Moon}}, \bibinfo {author} {\bibfnamefont {J.}~\bibnamefont {Yamaura}}, \bibinfo {author} {\bibfnamefont {Z.}~\bibnamefont {Hiroi}}, \emph {et~al.},\ }\href@noop {} {\bibfield  {journal} {\bibinfo  {journal} {Physical Review Letters}\ }\textbf {\bibinfo {volume} {118}},\ \bibinfo {pages} {117201} (\bibinfo {year} {2017})}\BibitemShut {NoStop}%
\bibitem [{\citenamefont {Son}\ \emph {et~al.}(2019)\citenamefont {Son}, \citenamefont {Park}, \citenamefont {Kim}, \citenamefont {Cho}, \citenamefont {Kim}, \citenamefont {Sandilands}, \citenamefont {Sohn}, \citenamefont {Park}, \citenamefont {Moon},\ and\ \citenamefont {Noh}}]{Son_YIO_2019}%
  \BibitemOpen
  \bibfield  {author} {\bibinfo {author} {\bibfnamefont {J.}~\bibnamefont {Son}}, \bibinfo {author} {\bibfnamefont {B.~C.}\ \bibnamefont {Park}}, \bibinfo {author} {\bibfnamefont {C.~H.}\ \bibnamefont {Kim}}, \bibinfo {author} {\bibfnamefont {H.}~\bibnamefont {Cho}}, \bibinfo {author} {\bibfnamefont {S.~Y.}\ \bibnamefont {Kim}}, \bibinfo {author} {\bibfnamefont {L.~J.}\ \bibnamefont {Sandilands}}, \bibinfo {author} {\bibfnamefont {C.}~\bibnamefont {Sohn}}, \bibinfo {author} {\bibfnamefont {J.-G.}\ \bibnamefont {Park}}, \bibinfo {author} {\bibfnamefont {S.~J.}\ \bibnamefont {Moon}},\ and\ \bibinfo {author} {\bibfnamefont {T.~W.}\ \bibnamefont {Noh}},\ }\href@noop {} {\bibfield  {journal} {\bibinfo  {journal} {npj Quantum materials}\ }\textbf {\bibinfo {volume} {4}},\ \bibinfo {pages} {17} (\bibinfo {year} {2019})}\BibitemShut {NoStop}%
\bibitem [{\citenamefont {Seth}\ \emph {et~al.}(2022)\citenamefont {Seth}, \citenamefont {Bhattacharjee},\ and\ \citenamefont {Moessner}}]{Seth2022}%
  \BibitemOpen
  \bibfield  {author} {\bibinfo {author} {\bibfnamefont {A.}~\bibnamefont {Seth}}, \bibinfo {author} {\bibfnamefont {S.}~\bibnamefont {Bhattacharjee}},\ and\ \bibinfo {author} {\bibfnamefont {R.}~\bibnamefont {Moessner}},\ }\href {https://doi.org/10.1103/PhysRevB.106.054507} {\bibfield  {journal} {\bibinfo  {journal} {Phys. Rev. B}\ }\textbf {\bibinfo {volume} {106}},\ \bibinfo {pages} {054507} (\bibinfo {year} {2022})}\BibitemShut {NoStop}%
\bibitem [{\citenamefont {Millican}\ \emph {et~al.}(2007)\citenamefont {Millican}, \citenamefont {Macaluso}, \citenamefont {Nakatsuji}, \citenamefont {Machida}, \citenamefont {Maeno},\ and\ \citenamefont {Chan}}]{Millican2007}%
  \BibitemOpen
  \bibfield  {author} {\bibinfo {author} {\bibfnamefont {J.~N.}\ \bibnamefont {Millican}}, \bibinfo {author} {\bibfnamefont {R.~T.}\ \bibnamefont {Macaluso}}, \bibinfo {author} {\bibfnamefont {S.}~\bibnamefont {Nakatsuji}}, \bibinfo {author} {\bibfnamefont {Y.}~\bibnamefont {Machida}}, \bibinfo {author} {\bibfnamefont {Y.}~\bibnamefont {Maeno}},\ and\ \bibinfo {author} {\bibfnamefont {J.~Y.}\ \bibnamefont {Chan}},\ }\href {https://doi.org/10.1016/j.materresbull.2006.08.011} {\bibfield  {journal} {\bibinfo  {journal} {Materials research bulletin}\ }\textbf {\bibinfo {volume} {42}},\ \bibinfo {pages} {928} (\bibinfo {year} {2007})}\BibitemShut {NoStop}%
\bibitem [{\citenamefont {Zoghlin}\ \emph {et~al.}(2021)\citenamefont {Zoghlin}, \citenamefont {Schmehr}, \citenamefont {Holgate}, \citenamefont {Dally}, \citenamefont {Liu}, \citenamefont {Laurita},\ and\ \citenamefont {Wilson}}]{Zoghlin2021}%
  \BibitemOpen
  \bibfield  {author} {\bibinfo {author} {\bibfnamefont {E.}~\bibnamefont {Zoghlin}}, \bibinfo {author} {\bibfnamefont {J.}~\bibnamefont {Schmehr}}, \bibinfo {author} {\bibfnamefont {C.}~\bibnamefont {Holgate}}, \bibinfo {author} {\bibfnamefont {R.}~\bibnamefont {Dally}}, \bibinfo {author} {\bibfnamefont {Y.}~\bibnamefont {Liu}}, \bibinfo {author} {\bibfnamefont {G.}~\bibnamefont {Laurita}},\ and\ \bibinfo {author} {\bibfnamefont {S.~D.}\ \bibnamefont {Wilson}},\ }\href {https://doi.org/10.1103/PhysRevMaterials.5.084403} {\bibfield  {journal} {\bibinfo  {journal} {Phys. Rev. Mater.}\ }\textbf {\bibinfo {volume} {5}},\ \bibinfo {pages} {084403} (\bibinfo {year} {2021})}\BibitemShut {NoStop}%
\bibitem [{\citenamefont {Cardona}\ and\ \citenamefont {G{\"u}ntherodt}(2000)}]{Cardona2000}%
  \BibitemOpen
  \bibfield  {author} {\bibinfo {author} {\bibfnamefont {M.}~\bibnamefont {Cardona}}\ and\ \bibinfo {author} {\bibfnamefont {G.}~\bibnamefont {G{\"u}ntherodt}},\ }\href@noop {} {\emph {\bibinfo {title} {Light Scattering in Solids VII: Crystal-Field and Magnetic Excitations}}},\ Vol.~\bibinfo {volume} {75}\ (\bibinfo  {publisher} {Springer},\ \bibinfo {year} {2000})\BibitemShut {NoStop}%
\bibitem [{\citenamefont {Xu}\ \emph {et~al.}(2022)\citenamefont {Xu}, \citenamefont {Man}, \citenamefont {Tang}, \citenamefont {Ohtsuki}, \citenamefont {Baidya}, \citenamefont {Nakatsuji}, \citenamefont {Vanderbilt},\ and\ \citenamefont {Drichko}}]{Xu2022phonons}%
  \BibitemOpen
  \bibfield  {author} {\bibinfo {author} {\bibfnamefont {Y.}~\bibnamefont {Xu}}, \bibinfo {author} {\bibfnamefont {H.}~\bibnamefont {Man}}, \bibinfo {author} {\bibfnamefont {N.}~\bibnamefont {Tang}}, \bibinfo {author} {\bibfnamefont {T.}~\bibnamefont {Ohtsuki}}, \bibinfo {author} {\bibfnamefont {S.}~\bibnamefont {Baidya}}, \bibinfo {author} {\bibfnamefont {S.}~\bibnamefont {Nakatsuji}}, \bibinfo {author} {\bibfnamefont {D.}~\bibnamefont {Vanderbilt}},\ and\ \bibinfo {author} {\bibfnamefont {N.}~\bibnamefont {Drichko}},\ }\href {https://doi.org/10.1103/PhysRevB.105.075137} {\bibfield  {journal} {\bibinfo  {journal} {Phys. Rev. B}\ }\textbf {\bibinfo {volume} {105}},\ \bibinfo {pages} {075137} (\bibinfo {year} {2022})}\BibitemShut {NoStop}%
\bibitem [{\citenamefont {Watahiki}\ \emph {et~al.}(2011)\citenamefont {Watahiki}, \citenamefont {Tomiyasu}, \citenamefont {Matsuhira}, \citenamefont {Iwasa}, \citenamefont {Yokoyama}, \citenamefont {Takagi}, \citenamefont {Wakeshima},\ and\ \citenamefont {Hinatsu}}]{Watahiki2011}%
  \BibitemOpen
  \bibfield  {author} {\bibinfo {author} {\bibfnamefont {M.}~\bibnamefont {Watahiki}}, \bibinfo {author} {\bibfnamefont {K.}~\bibnamefont {Tomiyasu}}, \bibinfo {author} {\bibfnamefont {K.}~\bibnamefont {Matsuhira}}, \bibinfo {author} {\bibfnamefont {K.}~\bibnamefont {Iwasa}}, \bibinfo {author} {\bibfnamefont {M.}~\bibnamefont {Yokoyama}}, \bibinfo {author} {\bibfnamefont {S.}~\bibnamefont {Takagi}}, \bibinfo {author} {\bibfnamefont {M.}~\bibnamefont {Wakeshima}},\ and\ \bibinfo {author} {\bibfnamefont {Y.}~\bibnamefont {Hinatsu}},\ }in\ \href {https://doi.org/10.1088/1742-6596/320/1/012080} {\emph {\bibinfo {booktitle} {Journal of Physics: Conference Series}}},\ Vol.\ \bibinfo {volume} {320}\ (\bibinfo {organization} {IOP Publishing},\ \bibinfo {year} {2011})\ p.\ \bibinfo {pages} {012080}\BibitemShut {NoStop}%
\bibitem [{\citenamefont {Xu}\ \emph {et~al.}(2015)\citenamefont {Xu}, \citenamefont {Anand}, \citenamefont {Bera}, \citenamefont {Frontzek}, \citenamefont {Abernathy}, \citenamefont {Casati}, \citenamefont {Siemensmeyer},\ and\ \citenamefont {Lake}}]{Xu2015}%
  \BibitemOpen
  \bibfield  {author} {\bibinfo {author} {\bibfnamefont {J.}~\bibnamefont {Xu}}, \bibinfo {author} {\bibfnamefont {V.~K.}\ \bibnamefont {Anand}}, \bibinfo {author} {\bibfnamefont {A.~K.}\ \bibnamefont {Bera}}, \bibinfo {author} {\bibfnamefont {M.}~\bibnamefont {Frontzek}}, \bibinfo {author} {\bibfnamefont {D.~L.}\ \bibnamefont {Abernathy}}, \bibinfo {author} {\bibfnamefont {N.}~\bibnamefont {Casati}}, \bibinfo {author} {\bibfnamefont {K.}~\bibnamefont {Siemensmeyer}},\ and\ \bibinfo {author} {\bibfnamefont {B.}~\bibnamefont {Lake}},\ }\href {https://doi.org/10.1103/PhysRevB.92.224430} {\bibfield  {journal} {\bibinfo  {journal} {Phys. Rev. B}\ }\textbf {\bibinfo {volume} {92}},\ \bibinfo {pages} {224430} (\bibinfo {year} {2015})}\BibitemShut {NoStop}%
\bibitem [{\citenamefont {Kim}\ \emph {et~al.}(2012)\citenamefont {Kim}, \citenamefont {Chen}, \citenamefont {Wang}, \citenamefont {Shi}, \citenamefont {Miotkowski}, \citenamefont {Chen}, \citenamefont {Sharma}, \citenamefont {Lima~Sharma}, \citenamefont {Hekmaty}, \citenamefont {Jiang} \emph {et~al.}}]{Kim2012}%
  \BibitemOpen
  \bibfield  {author} {\bibinfo {author} {\bibfnamefont {Y.}~\bibnamefont {Kim}}, \bibinfo {author} {\bibfnamefont {X.}~\bibnamefont {Chen}}, \bibinfo {author} {\bibfnamefont {Z.}~\bibnamefont {Wang}}, \bibinfo {author} {\bibfnamefont {J.}~\bibnamefont {Shi}}, \bibinfo {author} {\bibfnamefont {I.}~\bibnamefont {Miotkowski}}, \bibinfo {author} {\bibfnamefont {Y.}~\bibnamefont {Chen}}, \bibinfo {author} {\bibfnamefont {P.}~\bibnamefont {Sharma}}, \bibinfo {author} {\bibfnamefont {A.}~\bibnamefont {Lima~Sharma}}, \bibinfo {author} {\bibfnamefont {M.}~\bibnamefont {Hekmaty}}, \bibinfo {author} {\bibfnamefont {Z.}~\bibnamefont {Jiang}}, \emph {et~al.},\ }\href {https://doi.org/https://doi.org/10.1063/1.3685465} {\bibfield  {journal} {\bibinfo  {journal} {Applied Physics Letters}\ }\textbf {\bibinfo {volume} {100}},\ \bibinfo {pages} {071907} (\bibinfo {year} {2012})}\BibitemShut {NoStop}%
\bibitem [{\citenamefont {Osterhoudt}\ \emph {et~al.}(2021)\citenamefont {Osterhoudt}, \citenamefont {Wang}, \citenamefont {Garcia}, \citenamefont {Plisson}, \citenamefont {Gooth}, \citenamefont {Felser}, \citenamefont {Narang},\ and\ \citenamefont {Burch}}]{Osterhoudt2021}%
  \BibitemOpen
  \bibfield  {author} {\bibinfo {author} {\bibfnamefont {G.~B.}\ \bibnamefont {Osterhoudt}}, \bibinfo {author} {\bibfnamefont {Y.}~\bibnamefont {Wang}}, \bibinfo {author} {\bibfnamefont {C.~A.}\ \bibnamefont {Garcia}}, \bibinfo {author} {\bibfnamefont {V.~M.}\ \bibnamefont {Plisson}}, \bibinfo {author} {\bibfnamefont {J.}~\bibnamefont {Gooth}}, \bibinfo {author} {\bibfnamefont {C.}~\bibnamefont {Felser}}, \bibinfo {author} {\bibfnamefont {P.}~\bibnamefont {Narang}},\ and\ \bibinfo {author} {\bibfnamefont {K.~S.}\ \bibnamefont {Burch}},\ }\href@noop {} {\bibfield  {journal} {\bibinfo  {journal} {Physical Review X}\ }\textbf {\bibinfo {volume} {11}},\ \bibinfo {pages} {011017} (\bibinfo {year} {2021})}\BibitemShut {NoStop}%
\bibitem [{\citenamefont {Hatnean~Ciomaga}\ \emph {et~al.}(2015)\citenamefont {Hatnean~Ciomaga}, \citenamefont {Lees}, \citenamefont {Petrenko}, \citenamefont {Keeble}, \citenamefont {Balakrishnan}, \citenamefont {Gutmann}, \citenamefont {Klekovkina},\ and\ \citenamefont {Malkin}}]{Hatnean2015}%
  \BibitemOpen
  \bibfield  {author} {\bibinfo {author} {\bibfnamefont {M.}~\bibnamefont {Hatnean~Ciomaga}}, \bibinfo {author} {\bibfnamefont {M.~R.}\ \bibnamefont {Lees}}, \bibinfo {author} {\bibfnamefont {O.~A.}\ \bibnamefont {Petrenko}}, \bibinfo {author} {\bibfnamefont {D.~S.}\ \bibnamefont {Keeble}}, \bibinfo {author} {\bibfnamefont {G.}~\bibnamefont {Balakrishnan}}, \bibinfo {author} {\bibfnamefont {M.~J.}\ \bibnamefont {Gutmann}}, \bibinfo {author} {\bibfnamefont {V.~V.}\ \bibnamefont {Klekovkina}},\ and\ \bibinfo {author} {\bibfnamefont {B.~Z.}\ \bibnamefont {Malkin}},\ }\href {https://doi.org/10.1103/PhysRevB.91.174416} {\bibfield  {journal} {\bibinfo  {journal} {Phys. Rev. B}\ }\textbf {\bibinfo {volume} {91}},\ \bibinfo {pages} {174416} (\bibinfo {year} {2015})}\BibitemShut {NoStop}%
\bibitem [{\citenamefont {Ueda}\ \emph {et~al.}(2019)\citenamefont {Ueda}, \citenamefont {Kaneko}, \citenamefont {Subedi}, \citenamefont {Minola}, \citenamefont {Kim}, \citenamefont {Fujioka}, \citenamefont {Tokura},\ and\ \citenamefont {Keimer}}]{Ueda2019}%
  \BibitemOpen
  \bibfield  {author} {\bibinfo {author} {\bibfnamefont {K.}~\bibnamefont {Ueda}}, \bibinfo {author} {\bibfnamefont {R.}~\bibnamefont {Kaneko}}, \bibinfo {author} {\bibfnamefont {A.}~\bibnamefont {Subedi}}, \bibinfo {author} {\bibfnamefont {M.}~\bibnamefont {Minola}}, \bibinfo {author} {\bibfnamefont {B.~J.}\ \bibnamefont {Kim}}, \bibinfo {author} {\bibfnamefont {J.}~\bibnamefont {Fujioka}}, \bibinfo {author} {\bibfnamefont {Y.}~\bibnamefont {Tokura}},\ and\ \bibinfo {author} {\bibfnamefont {B.}~\bibnamefont {Keimer}},\ }\href {https://doi.org/10.1103/PhysRevB.100.115157} {\bibfield  {journal} {\bibinfo  {journal} {Phys. Rev. B}\ }\textbf {\bibinfo {volume} {100}},\ \bibinfo {pages} {115157} (\bibinfo {year} {2019})}\BibitemShut {NoStop}%
\bibitem [{\citenamefont {Rosalin}\ \emph {et~al.}(2023)\citenamefont {Rosalin}, \citenamefont {Telang}, \citenamefont {Singh}, \citenamefont {Muthu},\ and\ \citenamefont {Sood}}]{Rosalin_PIO_2023}%
  \BibitemOpen
  \bibfield  {author} {\bibinfo {author} {\bibfnamefont {M.}~\bibnamefont {Rosalin}}, \bibinfo {author} {\bibfnamefont {P.}~\bibnamefont {Telang}}, \bibinfo {author} {\bibfnamefont {S.}~\bibnamefont {Singh}}, \bibinfo {author} {\bibfnamefont {D.~V.~S.}\ \bibnamefont {Muthu}},\ and\ \bibinfo {author} {\bibfnamefont {A.~K.}\ \bibnamefont {Sood}},\ }\href {https://doi.org/10.1103/PhysRevB.108.195144} {\bibfield  {journal} {\bibinfo  {journal} {Phys. Rev. B}\ }\textbf {\bibinfo {volume} {108}},\ \bibinfo {pages} {195144} (\bibinfo {year} {2023})}\BibitemShut {NoStop}%
\bibitem [{\citenamefont {Rosalin}\ \emph {et~al.}(2024)\citenamefont {Rosalin}, \citenamefont {Telang}, \citenamefont {Singh}, \citenamefont {Muthu},\ and\ \citenamefont {Sood}}]{Rosalin_AIO_2024}%
  \BibitemOpen
  \bibfield  {author} {\bibinfo {author} {\bibfnamefont {M.}~\bibnamefont {Rosalin}}, \bibinfo {author} {\bibfnamefont {P.}~\bibnamefont {Telang}}, \bibinfo {author} {\bibfnamefont {S.}~\bibnamefont {Singh}}, \bibinfo {author} {\bibfnamefont {D.}~\bibnamefont {Muthu}},\ and\ \bibinfo {author} {\bibfnamefont {A.}~\bibnamefont {Sood}},\ }\href@noop {} {\bibfield  {journal} {\bibinfo  {journal} {Physical Review B}\ }\textbf {\bibinfo {volume} {109}},\ \bibinfo {pages} {184434} (\bibinfo {year} {2024})}\BibitemShut {NoStop}%
\bibitem [{\citenamefont {Wen}\ \emph {et~al.}(2017)\citenamefont {Wen}, \citenamefont {Koohpayeh}, \citenamefont {Ross}, \citenamefont {Trump}, \citenamefont {McQueen}, \citenamefont {Kimura}, \citenamefont {Nakatsuji}, \citenamefont {Qiu}, \citenamefont {Pajerowski}, \citenamefont {Copley},\ and\ \citenamefont {Broholm}}]{Wen2017}%
  \BibitemOpen
  \bibfield  {author} {\bibinfo {author} {\bibfnamefont {J.-J.}\ \bibnamefont {Wen}}, \bibinfo {author} {\bibfnamefont {S.~M.}\ \bibnamefont {Koohpayeh}}, \bibinfo {author} {\bibfnamefont {K.~A.}\ \bibnamefont {Ross}}, \bibinfo {author} {\bibfnamefont {B.~A.}\ \bibnamefont {Trump}}, \bibinfo {author} {\bibfnamefont {T.~M.}\ \bibnamefont {McQueen}}, \bibinfo {author} {\bibfnamefont {K.}~\bibnamefont {Kimura}}, \bibinfo {author} {\bibfnamefont {S.}~\bibnamefont {Nakatsuji}}, \bibinfo {author} {\bibfnamefont {Y.}~\bibnamefont {Qiu}}, \bibinfo {author} {\bibfnamefont {D.~M.}\ \bibnamefont {Pajerowski}}, \bibinfo {author} {\bibfnamefont {J.~R.~D.}\ \bibnamefont {Copley}},\ and\ \bibinfo {author} {\bibfnamefont {C.~L.}\ \bibnamefont {Broholm}},\ }\href {https://doi.org/10.1103/PhysRevLett.118.107206} {\bibfield  {journal} {\bibinfo  {journal} {Phys. Rev. Lett.}\ }\textbf {\bibinfo {volume} {118}},\ \bibinfo {pages} {107206} (\bibinfo {year} {2017})}\BibitemShut {NoStop}%
\end{thebibliography}%

\end{document}